\documentclass{aa}
\usepackage{graphicx}
\usepackage[varg]{txfonts}
\usepackage{cases}
\usepackage{natbib}

\begin{document}

   \title{Simultaneous transverse oscillations of a coronal loop and a filament excited by a circular-ribbon flare}

   \author{Q. M. Zhang\inst{1,2,3,4}}

   \institute{Key Laboratory of Dark Matter and Space Astronomy, Purple Mountain Observatory, CAS, Nanjing 210023, PR China \\
              \email{zhangqm@pmo.ac.cn}
              \and
              School of Astronomy and Space Science, Nanjing University, Nanjing 210023, PR China \\
              \and
              State Key Laboratory of Space Weather, Chinese Academy of Sciences, Beijing 100190, PR China \\
              \and
              State Key Laboratory of Lunar and Planetary Sciences, Macau University of Science and Technology, Macau, PR China \\
              }

   \date{Received; accepted}
    \titlerunning{Simultaneous transverse oscillations of a coronal loop and a filament}
    \authorrunning{Q. M. Zhang}

 \abstract
   {}
   {To investigate the excitation of kink oscillations in coronal loops and filaments, 
   a C3.4 circular-ribbon flare (CRF) associated with a blowout jet in active region 12434 on 2015 October 16 is analyzed.}
   {The flare was observed in ultraviolet (UV) and extreme-ultraviolet (EUV) wavelengths by the Atmospheric Imaging Assembly (AIA) 
   on board the Solar Dynamics Observatory (SDO) spacecraft.
   The line-of-sight (LOS) magnetograms of the photosphere were observed by the Helioseismic and Magnetic Imager (HMI) on board SDO.
   Soft X-ray (SXR) fluxes of the flares in 0.5$-$4 and 1$-$8 {\AA} were recorded by the GOES spacecraft.}
   {The flare excited small-amplitude kink oscillation of a remote coronal loop. The oscillation lasted for $\ge$4 cycles without significant damping. 
   The amplitude and period are 0.3$\pm$0.1 Mm and 207$\pm$12 s.
   Interestingly, the flare also excited transverse oscillation of a remote filament. The oscillation lasted for $\sim$3.5 cycles with decaying amplitudes. The initial amplitude is 1.7$-$2.2 Mm. 
   The period and damping time are 437$-$475 s and 1142$-$1600 s. 
   The starting times of simultaneous oscillations of coronal loop and filament were concurrent with the hard X-ray peak time.
   Though small in size and short in lifetime, the flare set off a chain reaction. It generated a bright secondary flare ribbon (SFR) in the chromosphere, 
   remote brightening (RB) that was cospatial with the filament, and intermittent, jet-like flow propagating in the northeast direction.}
   {The loop oscillation is most probably excited by the flare-induced blast wave at a speed of $\ge$1300 km s$^{-1}$.
   The excitation of the filament oscillation is more complicated. The blast wave triggers secondary magnetic reconnection far from the main flare,
   which not only heats the local plasma to higher temperatures (SFR and RB), but produces jet-like flow (i.e., reconnection outflow) as well. 
   The filament is disturbed by the secondary magnetic reconnection and experiences transverse oscillation.
   The findings give new insight into the excitation of transverse oscillations of coronal loops and filaments.}

 \keywords{Sun: flares -- Sun: filaments, prominences -- Sun: oscillations}

\maketitle

\section{Introduction} \label{s-intro}
Waves and oscillations are prevalent in the solar atmosphere \citep[see][and references therein]{oli02,naka05,arr12,de12}. 
After being disturbed by an external driver, a static coronal loop may deviate from its equilibrium position and experience oscillations \citep{edw83}. 
Transverse coronal loop oscillations are most frequently excited by adjacent solar flares \citep{asch99,naka99,wang04,zim15,li17,li20} and occasionally by shock waves \citep{hud04}
or coronal extreme-ultraviolet (EUV) waves \citep{kum13}. 
The initial amplitudes of loop displacement range from a few to 30 Mm and most of them are less than 10 Mm \citep{nech19}.
Two quantities, i.e. the loop length and density contrast play a relevant role in determining the amplitude of oscillation \citep{ter07}.
\citet{god16b} analyzed 58 kink oscillation events observed by the Atmospheric Imaging Assembly \citep[AIA;][]{lem12} on board the Solar Dynamics Observatory (SDO) 
during its first four years of operation. It is found that the period is proportional to the loop length \citep{god16a}.
The commonly observed transverse oscillations of standing kink mode provide a useful tool to infer the magnetic field strength and Alfv\'{e}n speed of the coronal loops, 
which are hard to measure in a direct way \citep{naka01,ver04,van08,wv12,yuan16,arr19}.
Some of the small-amplitude ($\la$0.5 Mm) oscillations hardly attenuate with time, which are called decayless oscillations \citep{anf13,anf15,li18a}.
\citet{nis13} reported small-amplitude, decayless loop oscillations before a flare and high-amplitude, decaying oscillations after the flare.

Circular-ribbon flares (CRFs) are a special type of flares, whose outer ribbons surrounding the compact inner ribbons show a circular or elliptical shape \citep{mas09,zqm16}. 
Like typical two-ribbon flares, CRFs are also capable of triggering transverse loop oscillations \citep{li18b}. \citet{zqm20b} investigated the transverse oscillations of an EUV loop 
excited by two homologous CRFs on 2014 March 5. The oscillations are divided into two stages in their development: 
the first-stage oscillation triggered by the C2.8 flare is decayless with lower amplitudes (310$-$510 km),
and the second-stage oscillation triggered by the M1.0 flare is decaying with larger amplitudes (1250$-$1280 km). 

Large-amplitude prominence or filament oscillations are divided into two categories according to their directions: transverse and longitudinal oscillations \citep{hyd66,ram66,kle69,tri09,luna18}.
Longitudinal oscillations can be triggered by microflares \citep{jing03,vrs07,zqm12,zqm17a}, flares \citep{li12,zqm20a}, coronal jets \citep{luna14}, shock waves \citep{shen14}, 
and failed filament eruptions \citep{maz20}. Transverse prominence oscillations are often excited by Moreton waves and/or EUV waves from a remote site of eruption at speeds of 
$\sim$1000 km s$^{-1}$ \citep[e.g.,][]{eto02,gil08,her11,asai12,dai12,gos12,liu12,shen17,zqm18}. 
Sometimes, they are triggered by magnetic reconnection as a result of magnetic flux emergence \citep{iso06,chen08}.
When an EUV jet from a remote AR arrives and collides with a filament, transverse oscillation of the filament may be generated \citep{zqm17b}.
Sophisticated numerical simulations have shed light on the triggering mechanism, restoring force, 
and damping mechanism of filament oscillations \citep[e.g.,][]{luna12,zqm13,zhou18,ad20,fan20,jel20,lia20}.
Several damping mechanisms have been proposed to interpret the observed attenuation of filament oscillations, 
such as thermal effects, resonant absorption in non-uniform media, and partial ionization effects \citep[see][and references therein]{arr11}.

So far, the excitation of kink oscillations in coronal loops and filaments is still controversial.
The motivation of this study is to investigate a C3.4 CRF in NOAA active region (AR) 12434, which excited transverse oscillations of a remote coronal loop and a remote filament on 2015 October 16.
This paper is organized as follows. Observations and results are presented in Sect.~\ref{s-res}. Possible origin of kink oscillations is discussed in Sect.~\ref{s-disc}. 
Finally, a brief summary is given in Sect.~\ref{s-sum}.

\begin{table}
\centering
\caption{Description of the observational parameters.}
\label{tab-1}
\begin{tabular}{cccc}
\hline\hline
Instrument & $\lambda$   &  Cad. & Pix. Size \\ 
                  & ({\AA})         &   (s)           & (\arcsec) \\
\hline
SDO/AIA & 171$-$335 &  12 & 0.6 \\
SDO/AIA & 1600        &  24 & 0.6 \\
SDO/HMI & 6173       & 45 & 0.6 \\
GOES     & 0.5$-$4.0 &  2.05 & ... \\
GOES     & 1$-$8    &  2.05 & ... \\
\hline
\end{tabular}
\end{table}

\section{Observations and Results} \label{s-res}

\subsection{Instruments and data analysis} \label{s-data}
The C3.4 flare was observed by SDO/AIA.
The AIA takes full-disk images in two ultraviolet (UV; 1600 and 1700 {\AA}) and seven EUV (94, 131, 171, 193, 211, 304, and 335 {\AA}) wavelengths.
The line-of-sight (LOS) magnetograms of the photosphere were observed by the Helioseismic and Magnetic Imager \citep[HMI;][]{sch12} on board SDO.
The level\_1 data of AIA and HMI were calibrated using the standard solar software (SSW) program \texttt{aia\_prep.pro} and \texttt{hmi\_prep.pro}, respectively.
Soft X-ray (SXR) light curves of the flare in 0.5$-$4 and 1$-$8 {\AA} were recorded by the GOES spacecraft. 
The observational parameters during 08:50$-$09:30 UT are listed in Table~\ref{tab-1}.

\subsection{Flare and blowout jet} \label{s-flare}
\begin{figure}
\includegraphics[width=8cm,clip=]{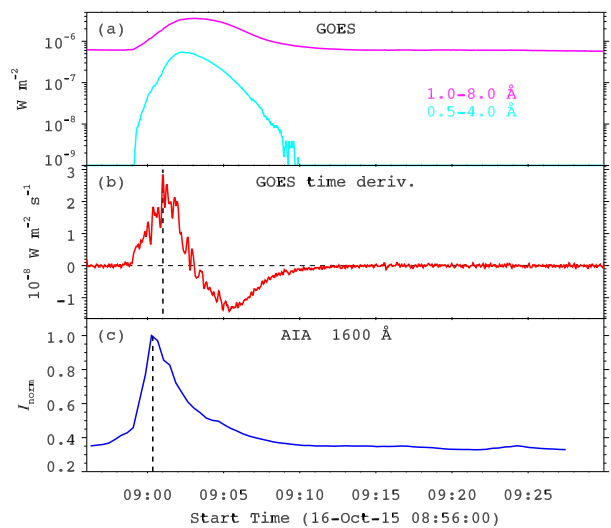}
\centering
\caption{(a) SXR light curves of the C3.4 flare in 0.5$-$4 {\AA} (cyan line) and 1$-$8 {\AA} (magenta line).
(b) Time derivative of the 1$-$8 {\AA} flux.
(c) Light curve of the flare in AIA 1600 {\AA}.}
\label{fig1}
\end{figure}

\begin{figure*}
\includegraphics[width=16cm,clip=]{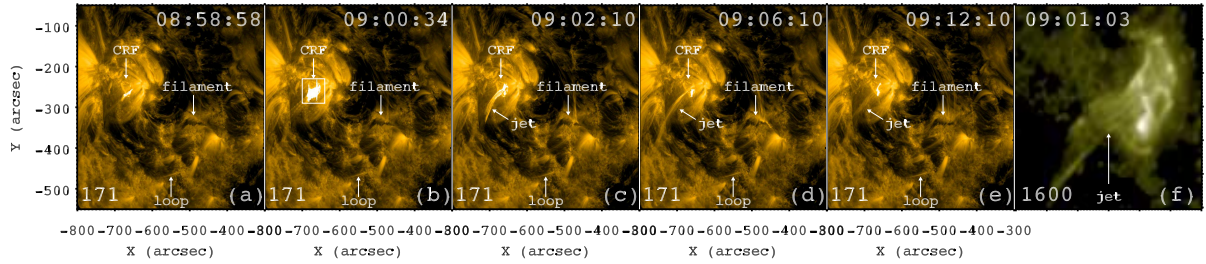}
\centering
\caption{(a-e) Snapshots of the AIA 171 {\AA} images. The arrows point to the CRF, bright loop, and dark filament. The white box in panel (b) signifies the flare region.
(f) Close-up of the flare region in AIA 1600 {\AA}. The whole evolution is shown in a movie (\textit{anim171.mov}) that is available online.}
\label{fig2}
\end{figure*}

In Fig.~\ref{fig1}, the top panel shows SXR light curves of the flare. The SXR emissions started to rise at $\sim$08:57 UT and reached peak values at $\sim$09:03 UT before declining gradually
until $\sim$09:12 UT. Therefore, the lifetime of the flare is $\sim$15 min, which is equal to that of the homologous C3.1 flare starting at $\sim$10:15 UT \citep{zqm16}. 
Figure~\ref{fig1}(b) shows the time derivative of the 1$-$8 {\AA} flux, which serves as a hard X-ray (HXR) proxy according 
to the Neupert effect. The HXR emission peaks at $\sim$09:01:00 UT. Light curve of the flare in AIA 1600 {\AA}, which is calculated by integrating the intensities of the flare region in Fig.~\ref{fig2}(f),
is plotted in Fig.~\ref{fig1}(c). The UV emission reaches its maximum $\sim$40 s before the HXR peak. Combining the UV and HXR light curves of the flare, it is concluded that the most impulsive 
release of energy occurred during 09:00$-$09:01 UT.

In Fig.~\ref{fig2}, the EUV images in 171 {\AA} demonstrate the whole evolution of the event. The EUV intensities of the flare started to increase at $\sim$08:58 UT and reached peak values at 
$\sim$09:00 UT (see panels (a) and (b)). The flare was accompanied by a curved blowout jet propagating in the southeast direction, which was observed in UV and EUV wavelengths (see panels (c-f)). 
The jet did not appear in the C2 white light (WL) coronagraph of the Large Angle Spectroscopic Coronagraph \citep[LASCO;][]{bru95} on board SOHO\footnote{http://cdaw.gsfc.nasa.gov/CME\_list/}, 
indicating that it did not evolve into a narrow coronal mass ejection (CME). Hence, the jet-related flare was a confined flare rather than an eruptive one.

\subsection{Coronal loop oscillation} \label{s-lo}
\begin{figure}
\includegraphics[width=8cm,clip=]{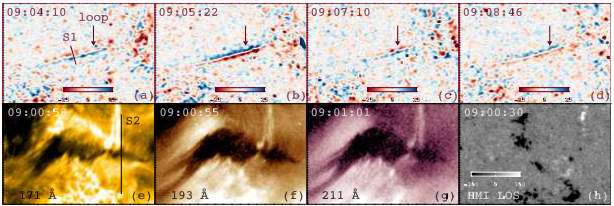}
\centering
\caption{(a-d) Running-difference images in 171 {\AA}, where blue (red) color represents intensity enhancement (weakening).
The slice (S1) is used to investigate the loop oscillation.
(e-g) Close-ups of the filament in 171, 193, and 211 {\AA}. The slice (S2) is used to investigate the filament oscillation.
(h) HMI LOS magnetogram associated with the filament. The field of view of each panel is 90$\arcsec\times$60$\arcsec$.}
\label{fig3}
\end{figure}

The flare excited transverse oscillation of a coronal loop, which is $\sim$230$\arcsec$ away from the flare site (see Fig.~\ref{fig2} and online movie \textit{anim171.mov}).
It should be emphasized that only the western part of oscillating loop close to the footpoint is clearly observed in EUV wavelengths. 
To better illustrate the displacements of the loop, the running difference technique is applied to the original EUV images.
Running-difference images in 171 {\AA} during 09:04$-$09:09 UT are displayed in the top panels of Fig.~\ref{fig3}. The blue (red) color represents intensity enhancement (weakening), respectively.
It is clear that the loop segment oscillated back and forth in a coherent way. Since the remaining segment of the loop is not visible, whether the kink oscillation is fundamental or harmonic is unknown.

\begin{figure}
\includegraphics[width=8cm,clip=]{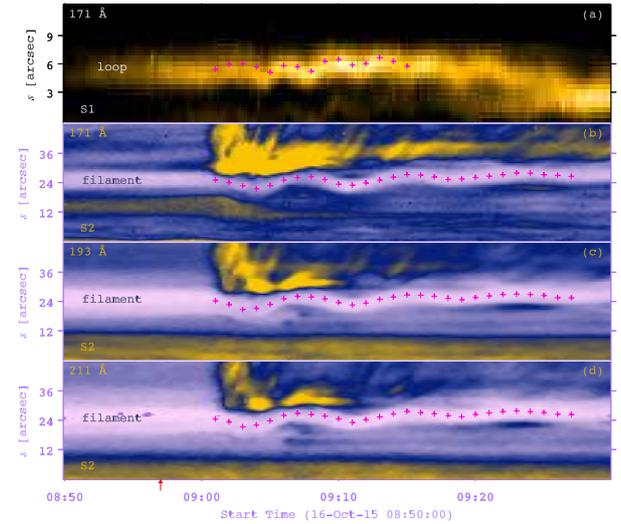}
\centering
\caption{(a) Time-distance diagram of S1 in 171 {\AA}, showing the transverse oscillation of the coronal loop.
The magenta plus symbols denote the central positions of the loop.
On the $y$-axis, $s=0$ and $s=12\arcsec$ denote the south and north endpoints of S1, respectively.
(b-d) Time-distance diagrams of S2 in different wavelengths, showing the transverse oscillation of the filament.
The magenta plus symbols denote the central positions of the filament.
On the $y$-axis, $s=0$ and $s=48\arcsec$ denote the south and north endpoints of S2, respectively.
The red arrow on the $x$-axis signifies the start time of C3.4 flare in SXR.}
\label{fig4}
\end{figure}

The loop oscillation is most definitely recognized in 171 {\AA}. To quantify the characteristics of the oscillation, an artificial slice (S1) with a length of 12$\arcsec$ across the loop is selected 
(see Fig.~\ref{fig3}(a)). Time-distance diagram of S1 in 171 {\AA} is shown in Fig.~\ref{fig4}(a). The magenta plus symbols denote the central positions of the loop.
The small-amplitude oscillation commenced at $\sim$09:01 UT and lasted for several cycles without significant damping.
The loop also experienced slight linear drifting motion at a speed of $\sim$0.6 km s$^{-1}$, which is listed in the second column of Table~\ref{tab-2}.

\begin{figure}
\includegraphics[width=8cm,clip=]{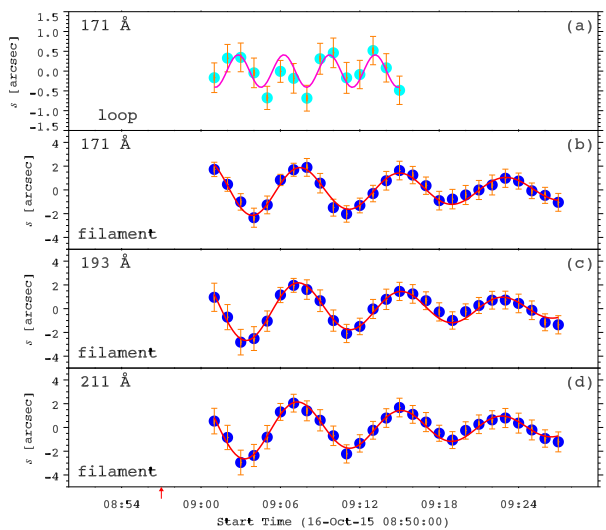}
\centering
\caption{(a) Detrended central positions of the coronal loop in 171 {\AA}.
Kink oscillation of the loop is fitted with a decayless (magenta line) sine function.
(b-d) Detrended central positions of the filament in 171, 193, and 211 {\AA}.
Transverse oscillation of the filament is fitted with an exponentially decaying (red lines) sine function.}
\label{fig5}
\end{figure}

The detrended central positions of the loop is drawn with cyan circles in Fig.~\ref{fig5}(a). 
Likewise, it is fitted with a decayless sine function using the standard SSW program \texttt{mpfit.pro} \citep{zqm20b}:
\begin{equation} \label{eqn-1}
  y_{1}=A_{1}\sin(\frac{2\pi}{P_1}t+\phi_1),
\end{equation}
where $A_{1}$ and $\phi_1$ stand for the initial amplitude of displacement and phase, $P_1$ denotes the period.
The derived values of $A_{1}=0.3\pm0.1$ Mm and $P_1=207\pm12$ s are listed in the third and fourth columns of Table~\ref{tab-2}.
In order to estimate the uncertainties of $A_{1}$ and $P_1$, one hundred Markov chain Monte Carlo \citep[MCMC;][]{sha17} sampling of $y_{1}$ is computed.
For each MC simulation, $y_{1}$ is pertubed by a small amount, which is randomly drawn from normal distribution with 
1$\sigma$ equals to the uncertainty of $y_1$. The curve fitting using \texttt{mpfit.pro} is repeated for each of 100 MC realizations \citep{cx12}.
The uncertainties of $A_{1}$ and $P_1$ are given at the 95\% credible interval \citep{arr19}.

The amplitude of loop oscillation in this study is slightly lower than that of the EUV loop during its decayless oscillation on 2014 March 5 \citep{zqm20b}.
\citet{ter07} examined the energy that an initial disturbance stories in the eigenmodes of coronal loops. 
It is found that the trapped energy in coronal loops decreases quickly with the distance of the pertubation,
which can explain the small amplitude of the oscillating loop that is $\sim$230$\arcsec$ away from the flare region.

\citet{ter05} investigated the excitation of trapped and leaky modes in coronal slabs. The leaky modes are characterized by short period oscillations and fast attenuations.
Trapped modes, on the contrary, have much longer periods and marginal attenuations. The observed decayless kink oscillation of the coronal loop in Fig.~\ref{fig4}(a) belong to the trapped mode.
The leaky mode before the trapped mode was absent, owing to the short period and damping time \citep{ter05}.

\subsection{Filament oscillation} \label{s-fo}
Interestingly, the flare not only excited kink oscillation of the coronal loop as described above, but also triggered transverse oscillation of the remote filament in the quiet region
(see Fig.~\ref{fig2} and online movie \textit{anim171.mov}). Close-ups of the filament observed at $\sim$09:01 UT in 171, 193, and 211 {\AA} are depicted in Fig.~\ref{fig3}(e-g).
The filament, consisting of fine dark threads, shows an $\mathcal{L}$-shape. 
The LOS magnetogram associated with the filament is displayed in Fig.~\ref{fig3}(h), where negative magnetic polarities dominate.
To investigate the filament oscillation, an artificial slice (S2) crossing the filament with a length of 48$\arcsec$ is selected in Fig.~\ref{fig3}(e). 
Time-distance diagrams of S2 in different wavelengths are displayed in Fig.~\ref{fig4}(b-d), where magenta plus symbols represent the central positions of the filament during the oscillation.
The oscillation started at $\sim$09:01 UT and lasted for a few cycles with damping amplitudes.
Linear drifting motion of the filament at a speed of $\sim$1.6 km s$^{-1}$ is extracted from the diagrams (see the second column of Table~\ref{tab-2}). 

To derive parameters of the kink oscillation, detrended central positions of the filament in different wavelengths are plotted with blue circles in Fig.~\ref{fig5}(b-d).
They are fitted with an exponentially decaying sine function \citep{zqm20b}:
\begin{equation} \label{eqn-2}
  y_{2}=A_{2}\sin(\frac{2\pi}{P_2}t+\phi_2)e^{-t/\tau_{2}},
\end{equation}
where $A_{2}$ and $\phi_2$ stand for the initial amplitude of displacement and phase, $P_2$ and $\tau_{2}$ signify the period and damping time.
The fitted values of $A_{2}$, $P_2$, $\tau_{2}$, and $\frac{\tau_{2}}{P_{2}}$ are listed in the middle four columns of Table~\ref{tab-2}.
The corresponding uncertainties of these parameters are obtained using the same method of MC simulation as described in Sect.~\ref{s-lo}.

\begin{table*}
\centering
\caption{Parameters of transverse loop oscillation (LO) and filament oscillation (FO) observed by AIA in different wavelengths. The quantity $b$ stands for the linear drift speed.}
\label{tab-2}
\begin{tabular}{c|cccccc}
\hline\hline
$\lambda$ & $b$ & $A$ & $P$ & $\tau$ & $\tau/P$ & Type \\
({\AA})     &  (km s$^{-1}$)  & (Mm) & (s) & (s) & &  \\
\hline
171 & 0.6 &  0.3$\pm$0.1 & 207$\pm$12 & ... & ... & LO \\
\hline
171 & 1.6 & 1.7$\pm$0.3 & 437$\pm$16 & 1600$\pm$758 & 3.6$\pm$1.8 & FO \\
193 & 1.6 & 2.2$\pm$0.5 & 471$\pm$18 & 1142$\pm$536 & 2.4$\pm$1.1 & FO \\
211 & 1.6 & 2.2$\pm$0.5 & 475$\pm$18 & 1151$\pm$530 & 2.4$\pm$1.1 & FO \\
\hline
\end{tabular}
\end{table*}

\section{Discussion} \label{s-disc}
Transverse oscillations are ubiquitous in the solar atmosphere, especially in coronal loops and prominences \citep{rud09}. 
\citet{zim15} analyzed 58 kink-oscillation events observed by SDO/AIA. 
The main conclusion of their work is that in nearly all events the excitation of kink oscillations is caused by the displacement of the loops from their equilibrium state 
as a result of a nearby lower coronal eruption/ejection and subsequent oscillatory relaxation of the loops. On the contrary, blast wave plays a minor role in the excitation.
In this study, the start time of loop oscillation is concurrent with the HXR peak time of the flare (see Fig.~\ref{fig1} and Fig.~\ref{fig5}). 
It is noted that the start times of both decayless and decaying loop oscillations were also coincident with the HXR peaks on 2014 March 5 \citep{zqm20b}, 
suggesting that the excitation of kink oscillations is related to the most impulsive release of magnetic energy rather than the very beginning of energy release. 
Hence, the real speeds of driver can potentially reach $\sim$1000 km s$^{-1}$, the typical speed of a blast wave \citep{toth11}.

\begin{figure*}
\includegraphics[width=16cm,clip=]{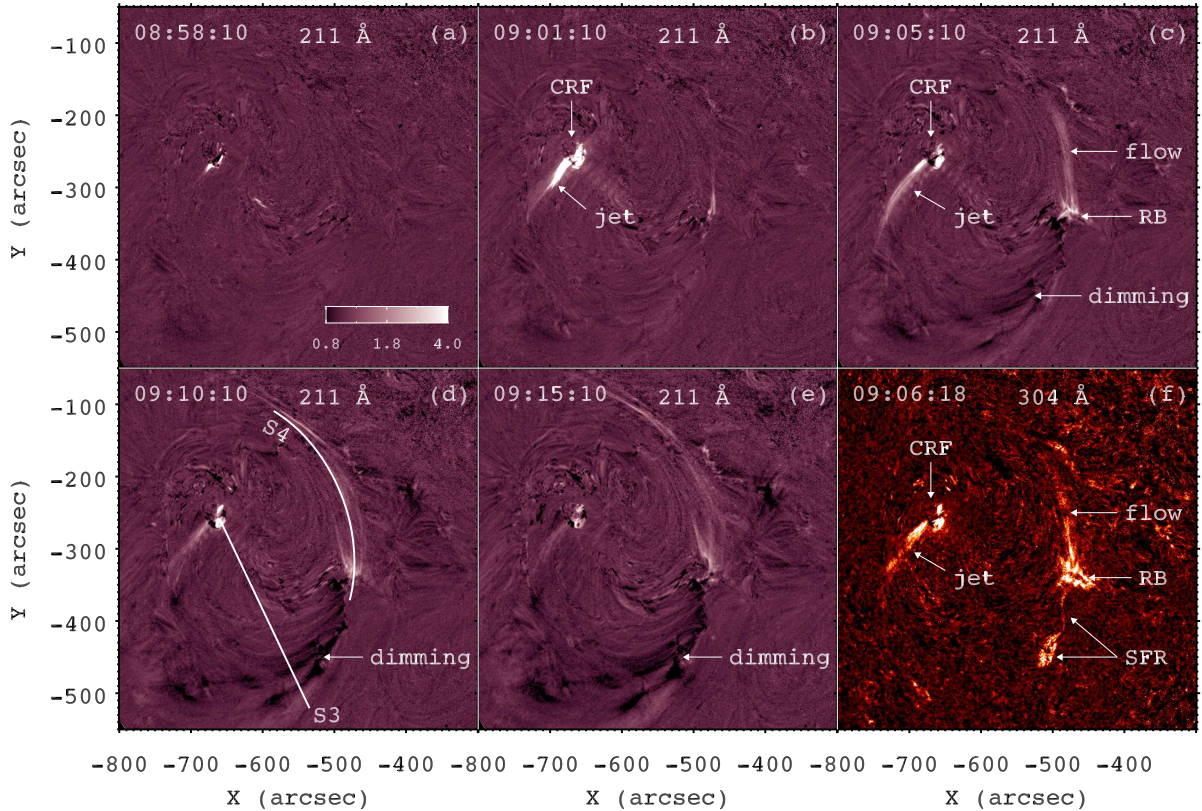}
\centering
\caption{Base-ratio images in 211 {\AA} (a-e) and 304 {\AA} (f). 
The white arrows point to the CRF, blowout jet, dark dimming, secondary flare ribbon (SFR), remote brightening (RB), and jet-like flow.
In panel (d), S3 is used to measure the speeds of a plausible blast wave. 
S4 is used to calculate the speeds of jet-like flow propagating in the northeast direction.
The whole evolution is shown in a movie (\textit{anim211.mov}) that is available online.}
\label{fig6}
\end{figure*}

\begin{figure}
\includegraphics[width=8cm,clip=]{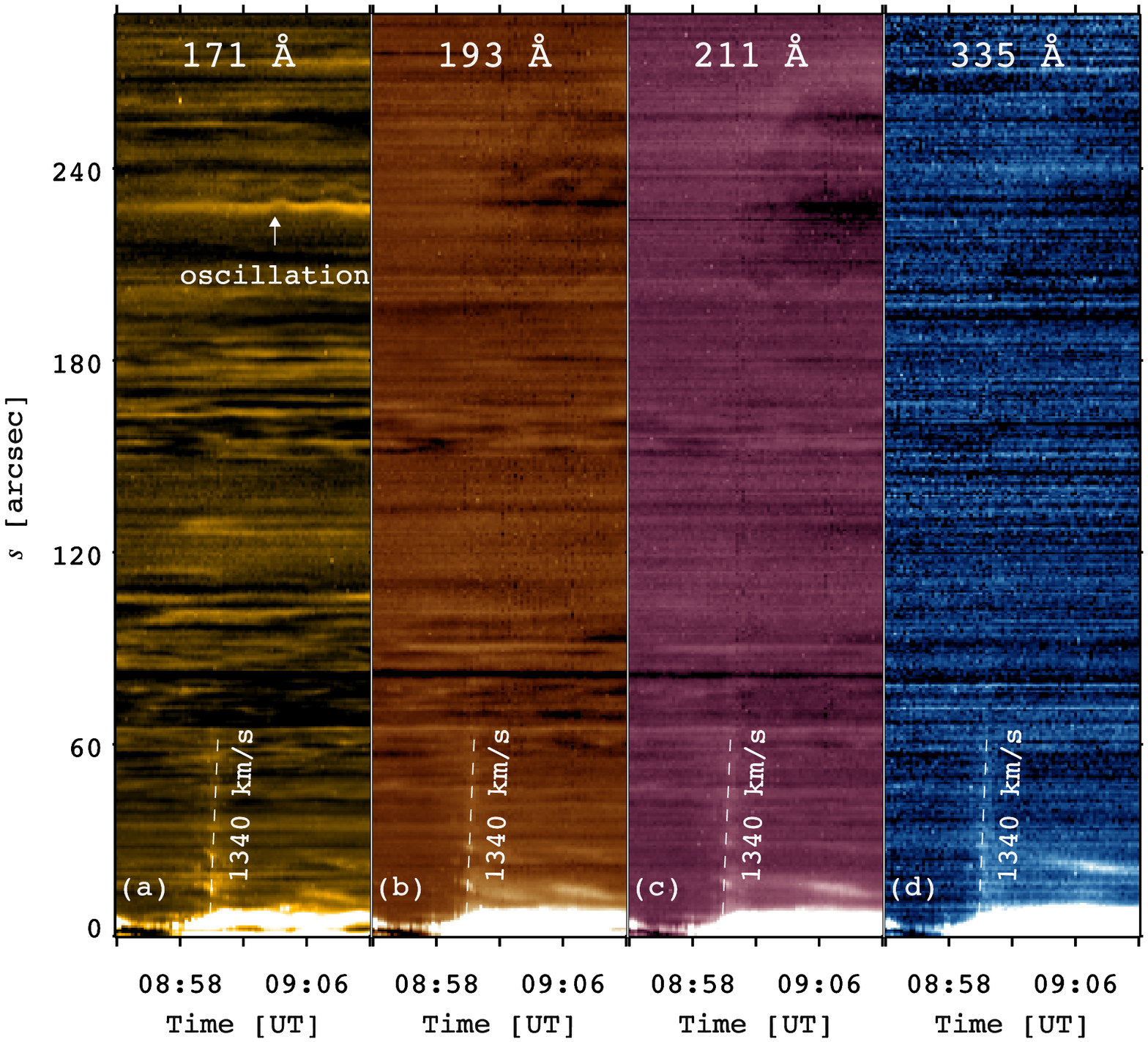}
\centering
\caption{Time-distance diagrams of S3 in different EUV wavelengths. The apparent speeds of the plausible blast wave are labeled.
On the $y$-axis, $s=0$ and $s=288\arcsec$ denote the flare site and southwest endpoint of S3.}
\label{fig7}
\end{figure}

From the online movie (\textit{anim171.mov}), a plausible signature of blast wave is crudely identified by eye. Figure~\ref{fig6} shows snapshots of the base-ratio images in 211 and 304 {\AA}.
The jet-related CRF, though small in size and short in lifetime, sets off a chain reaction. It generates large-scale remote dimming, a bright secondary flare ribbon (SFR) in the chromosphere, 
remote brightening (RB) that is cospatial with the filament, and jet-like flow propagating in the northeast direction.
In Fig.~\ref{fig6}(d), a long artificial slice (S3), starting from the flare site and passing through the oscillating loop, is selected to investigate the plausible blast wave.
Time-distance diagrams of S3 in different EUV wavelenths are plotted in Fig.~\ref{fig7}. A sharp linear structure, coincident with the UV and HXR peaks during 09:00$-$09:01 UT, 
is evident in all wavelengths. Using the slopes of the structure, the apparent speeds are calculated to be $\sim$1340 km s$^{-1}$, 
which has the same order of magnitude as blast waves in the corona \citep{hud04,toth11}.
The kink oscillation of the loop commences after the blast wave arrives. Therefore, the driver of loop oscillation is most probably the flare-induced blast wave.

\begin{figure}
\includegraphics[width=8cm,clip=]{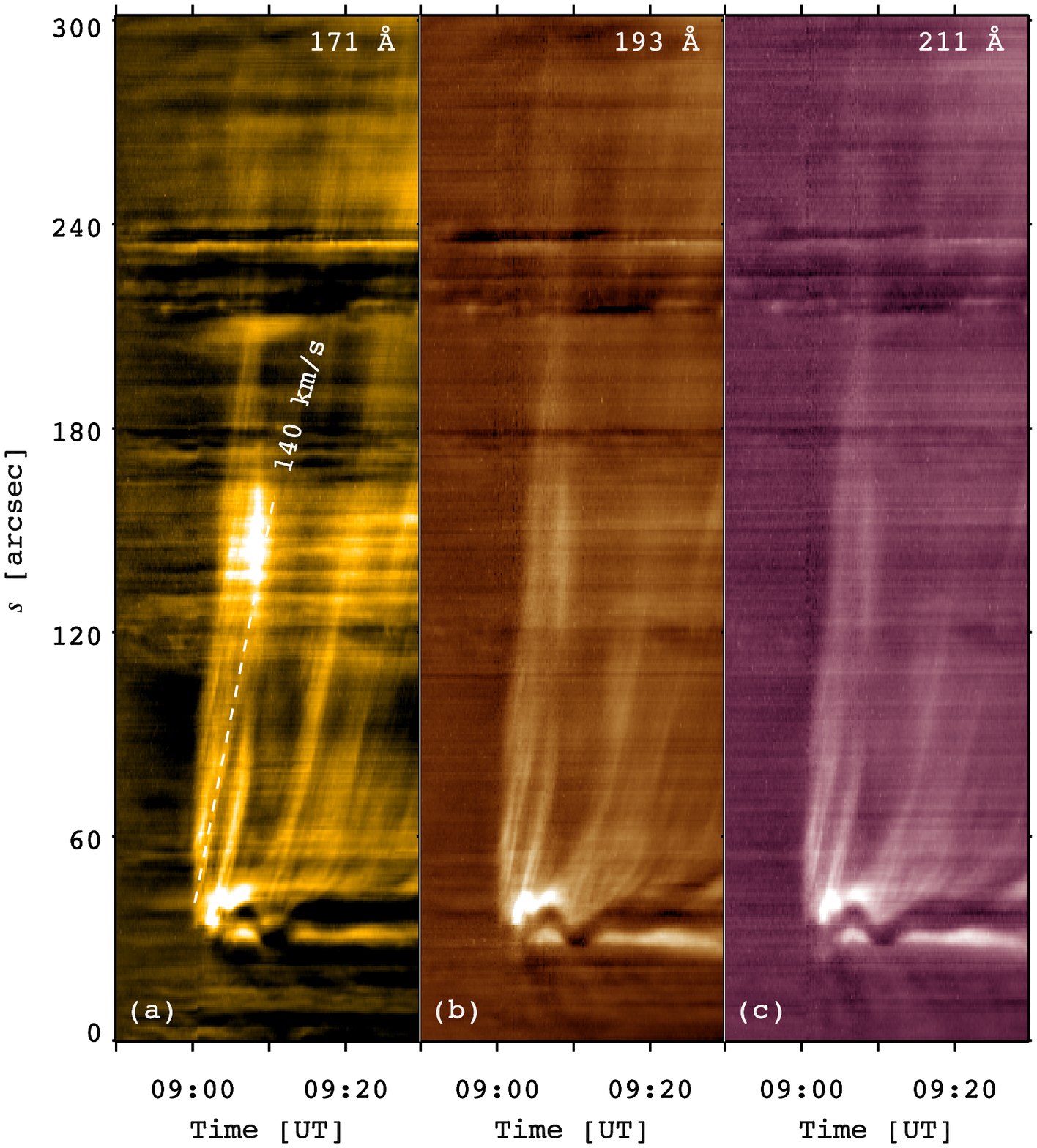}
\centering
\caption{Time-distance diagrams of S4 in 171, 193 and 211 {\AA}.
The apparent speed ($\sim$140 km s$^{-1}$) of the jet-like flow is labeled.
On the $y$-axis, $s=0$ and $s=301\arcsec$ denote the south and north endpoints of S4.}
\label{fig8}
\end{figure}

\citet{zj14} investigated the flare ribbons of 19 X-class flares observed by SDO/AIA, finding that 11 of the 16 well-detected events show multiple ribbons. 
The authors divided the ribbons into two types: normal ribbons that are connected by the post-flare loops (PFLs) and SFRs. The short-period SFRs, usually weaker than the primary ribbons, 
are not connected by PFLs. It is suggested that the magnetic reconnection associated with the SFRs is probably triggered by the flare-induced blast wave.
In this study of the C3.4 flare, it generates a SFR observed in 304 {\AA} (see Fig.~\ref{fig6}(f)). Meanwhile, the RB cospatial with the filament and jet-like flow propagating in the northeast direction 
are observed by AIA in EUV wavelengths, implying their multithermal nature. In Fig.~\ref{fig6}(d), a curved slice (S4) is selected to investigate the plasma flow. 
Time-slice diagrams of S4 in different EUV wavelengths are displayed in Fig.~\ref{fig8}. 
Intermittent plasma flow, originating from the RB and propagating at speeds of $\sim$140 km s$^{-1}$, is clearly demonstrated. 
The starting times of filament oscillation and jet-like flow are consistent with each other. 

\begin{figure}
\includegraphics[width=8cm,clip=]{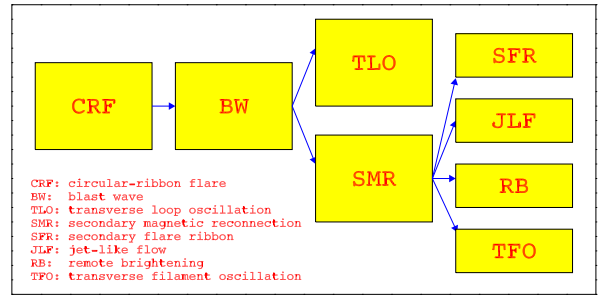}
\centering
\caption{Timeline of the whole event to illustrate a plausible scenario of transverse oscillations in a remote coronal loop and a remote filament excited by the CRF.}
\label{fig9}
\end{figure}

Inspired by the conjecture of \citet{zj14}, it is proposed that the flare-induced blast wave causes secondary magnetic reconnection far from the primary flare,
which not only heats the local plasma to higher temperatures (SFR and RB), but produces jet-like flow (i.e., reconnection outflow) as well. 
Meanwhile, the filament is disturbed by the magnetic reconnection and experiences transverse oscillation. 
The oscillating threads with the highest amplitude and longest cycles are excellently cospatial with the source of plasma flow (see Fig.~\ref{fig3}(e)).
The remaining part of the filament oscillated in the same direction with much faster damping and shorter lifetime (1$-$2 cycles).
Hence, the excitation of filament oscillation in this event is much more complicated than the situation in previously reported events where filament oscillations 
are directly excited when fast coronal EUV waves or Moreton waves impact the filaments from aside \citep[e.g.,][]{eto02,her11,dai12,gos12,liu12,shen17}.
The plausible blast wave is not a shock wave, since the type \textrm{II} radio burst associated with the shock wave was not observed.
It is noted that the scenario of kink oscillations triggered by a blast wave is a conjecture (see Fig.~\ref{fig9}), which needs to be validated in the future.
The oscillations of coronal loop and filament could not be triggered by the blowout jet, since the jet propagated in the southeast direction and covered a limited distance.

\section{Summary} \label{s-sum}
In this work, a C3.4 CRF associated with a blowout jet in AR 12434 on 2015 October 16 is studied. The main results are summarized as follows:

\begin{enumerate}
\item{The flare excited small-amplitude kink oscillation of a remote coronal loop. The oscillation lasted for $\ge$4 cycles without significant damping. 
The amplitude and period are 0.3$\pm$0.1 Mm and 207$\pm$12 s, respectively.
Intriguingly, the flare also excited transverse oscillation of a remote filament. The oscillation lasted for $\sim$3.5 cycles with decaying amplitudes. The initial amplitude is 1.7$-$2.2 Mm. 
The period and damping time are 437$-$475 s and 1142$-$1600 s. 
The starting times of simultaneous oscillations of the coronal loop and filament were concurrent with the HXR peak time.}
\item{Though small in size and short in lifetime, the flare set off a chain reaction. It generated a bright SFR in the chromosphere, RB that was cospatial with the filament, 
and intermittent, jet-like flow propagating in the northeast direction. The loop oscillation is most probably excited by the flare-induced blast wave at a speed of $\ge$1300 km s$^{-1}$.
The excitation of the filament oscillation is more complicated. The blast wave triggers secondary magnetic reconnection far from the primary flare,
which not only heats the local plasma to higher temperatures (SFR and RB), but produces intermittent, jet-like flow (i.e., reconnection outflow) as well. 
The filament is disturbed by the secondary magnetic reconnection and experiences transverse oscillation.
The results are important to understand the excitation of transverse oscillations of coronal loops and filaments.}
\end{enumerate}

\begin{acknowledgements}
The author is thankful for valuable comments and suggestions from the referee. 
The author appreciates Prof. Jun Zhang and Prof. Song Feng for inspiring discussion.
SDO is a mission of NASA\rq{}s Living With a Star Program. AIA and HMI data are courtesy of the NASA/SDO science teams.
This work is funded by NSFC grants (No. 11773079, 11790302), the International Cooperation and Interchange Program (11961131002), 
the Youth Innovation Promotion Association CAS, the Science and Technology Development Fund of Macau (275/2017/A), 
CAS Key Laboratory of Solar Activity, National Astronomical Observatories (KLSA202006),
and the project supported by the Specialized Research Fund for State Key Laboratories.
\end{acknowledgements}


\begin{thebibliography}{}
\bibitem[Adrover-Gonz{\'a}lez \& Terradas(2020)]{ad20} Adrover-Gonz{\'a}lez, A., \& Terradas, J.\ 2020, \aap, 633, A113
\bibitem[Anfinogentov et al.(2013)]{anf13} Anfinogentov, S., Nistic{\`o}, G., \& Nakariakov, V.~M.\ 2013, \aap, 560, A107
\bibitem[Anfinogentov et al.(2015)]{anf15} Anfinogentov, S.~A., Nakariakov, V.~M., \& Nistic{\`o}, G.\ 2015, \aap, 583, A136
\bibitem[Arregui \& Ballester(2011)]{arr11} Arregui, I., \& Ballester, J.~L.\ 2011, \ssr, 158, 169
\bibitem[Arregui et al.(2012)]{arr12} Arregui, I., Oliver, R., \& Ballester, J.~L.\ 2012, Living Reviews in Solar Physics, 9, 2
\bibitem[Arregui et al.(2019)]{arr19} Arregui, I., Montes-Sol{\'\i}s, M., \& Asensio Ramos, A.\ 2019, \aap, 625, A35
\bibitem[Asai et al.(2012)]{asai12} Asai, A., Ishii, T.~T., Isobe, H., et al.\ 2012, \apjl, 745, L18
\bibitem[Aschwanden et al.(1999)]{asch99} Aschwanden, M.~J., Fletcher, L., Schrijver, C.~J., et al.\ 1999, \apj, 520, 880
\bibitem[Brueckner et al.(1995)]{bru95} Brueckner, G.~E., Howard, R.~A., Koomen, M.~J., et al.\ 1995, \solphys, 162, 357
\bibitem[Chen et al.(2008)]{chen08} Chen, P.~F., Innes, D.~E., \& Solanki, S.~K.\ 2008, \aap, 484, 487
\bibitem[Cheng et al.(2012)]{cx12} Cheng, X., Zhang, J., Saar, S.~H., et al.\ 2012, \apj, 761, 62
\bibitem[Dai et al.(2012)]{dai12} Dai, Y., Ding, M.~D., Chen, P.~F., et al.\ 2012, \apj, 759, 55
\bibitem[De Moortel \& Nakariakov(2012)]{de12} De Moortel, I. \& Nakariakov, V.~M.\ 2012, Philosophical Transactions of the Royal Society of London Series A, 370, 3193
\bibitem[Edwin \& Roberts(1983)]{edw83} Edwin, P.~M., \& Roberts, B.\ 1983, \solphys, 88, 179
\bibitem[Eto et al.(2002)]{eto02} Eto, S., Isobe, H., Narukage, N., et al.\ 2002, \pasj, 54, 481
\bibitem[Fan(2020)]{fan20} Fan, Y.\ 2020, \apj, 898, 34
\bibitem[Gilbert et al.(2008)]{gil08} Gilbert, H.~R., Daou, A.~G., Young, D., Tripathi, D., \& Alexander, D.\ 2008, \apj, 685, 629
\bibitem[Goddard \& Nakariakov(2016)]{god16a} Goddard, C.~R., \& Nakariakov, V.~M.\ 2016, \aap, 590, L5
\bibitem[Goddard et al.(2016)]{god16b} Goddard, C.~R., Nistic{\`o}, G., Nakariakov, V.~M., et al.\ 2016, \aap, 585, A137
\bibitem[Gosain \& Foullon(2012)]{gos12} Gosain, S., \& Foullon, C.\ 2012, \apj, 761, 103
\bibitem[Hershaw et al.(2011)]{her11} Hershaw, J., Foullon, C., Nakariakov, V.~M., et al.\ 2011, \aap, 531, A53
\bibitem[Hudson \& Warmuth(2004)]{hud04} Hudson, H.~S., \& Warmuth, A.\ 2004, \apjl, 614, L85
\bibitem[Hyder(1966)]{hyd66} Hyder, C.~L.\ 1966, \zap, 63, 78
\bibitem[Isobe \& Tripathi(2006)]{iso06} Isobe, H., \& Tripathi, D.\ 2006, \aap, 449, L17
\bibitem[Jel{\'\i}nek et al.(2020)]{jel20} Jel{\'\i}nek, P., Karlick{\'y}, M., Smirnova, V.~V., et al.\ 2020, \aap, 637, A42
\bibitem[Jing et al.(2003)]{jing03} Jing, J., Lee, J., Spirock, T.~J., et al.\ 2003, \apjl, 584, L103
\bibitem[Kleczek \& Kuperus(1969)]{kle69} Kleczek, J., \& Kuperus, M.\ 1969, \solphys, 6, 72
\bibitem[Kumar et al.(2013)]{kum13} Kumar, P., Cho, K.-S., Chen, P.~F., et al.\ 2013, \solphys, 282, 523
\bibitem[Lemen et al.(2012)]{lem12} Lemen, J.~R., Title, A.~M., Akin, D.~J., et al.\ 2012, \solphys, 275, 17
\bibitem[Li \& Zhang(2012)]{li12} Li, T., \& Zhang, J.\ 2012, \apjl, 760, L10
\bibitem[Li et al.(2017)]{li17} Li, D., Ning, Z.~J., Huang, Y., et al.\ 2017, \apj, 849, 113
\bibitem[Li et al.(2018a)]{li18a} Li, D., Yuan, D., Su, Y.~N., et al.\ 2018a, \aap, 617, A86
\bibitem[Li et al.(2018b)]{li18b} Li, T., Yang, S., Zhang, Q., et al.\ 2018b, \apj, 859, 122
\bibitem[Li et al.(2020)]{li20} Li, D., Li, Y., Lu, L., et al.\ 2020, \apjl, 893, L17
\bibitem[Liakh et al.(2020)]{lia20} Liakh, V., Luna, M., \& Khomenko, E.\ 2020, \aap, 637, A75
\bibitem[Liu et al.(2012)]{liu12} Liu, W., Ofman, L., Nitta, N.~V., et al.\ 2012, \apj, 753, 52
\bibitem[Luna \& Karpen(2012)]{luna12} Luna, M., \& Karpen, J.\ 2012, \apjl, 750, L1
\bibitem[Luna et al.(2014)]{luna14} Luna, M., Knizhnik, K., Muglach, K., et al.\ 2014, \apj, 785, 79
\bibitem[Luna et al.(2018)]{luna18} Luna, M., Karpen, J., Ballester, J.~L., et al.\ 2018, \apjs, 236, 35
\bibitem[Masson et al.(2009)]{mas09} Masson, S., Pariat, E., Aulanier, G., et al.\ 2009, \apj, 700, 559
\bibitem[Mazumder et al.(2020)]{maz20} Mazumder, R., Pant, V., Luna, M., et al.\ 2020, \aap, 633, A12
\bibitem[Nakariakov et al.(1999)]{naka99} Nakariakov, V.~M., Ofman, L., Deluca, E.~E., et al.\ 1999, Science, 285, 862
\bibitem[Nakariakov \& Ofman(2001)]{naka01} Nakariakov, V.~M., \& Ofman, L.\ 2001, \aap, 372, L53
\bibitem[Nakariakov \& Verwichte(2005)]{naka05} Nakariakov, V.~M., \& Verwichte, E.\ 2005, Living Reviews in Solar Physics, 2, 3
\bibitem[Nechaeva et al.(2019)]{nech19} Nechaeva, A., Zimovets, I.~V., Nakariakov, V.~M., et al.\ 2019, \apjs, 241, 31
\bibitem[Nistic{\`o} et al.(2013)]{nis13} Nistic{\`o}, G., Nakariakov, V.~M., \& Verwichte, E.\ 2013, \aap, 552, A57
\bibitem[Oliver \& Ballester(2002)]{oli02} Oliver, R., \& Ballester, J.~L.\ 2002, \solphys, 206, 45
\bibitem[Ramsey \& Smith(1966)]{ram66} Ramsey, H.~E., \& Smith, S.~F.\ 1966, \aj, 71, 197
\bibitem[Ruderman \& Erd{\'e}lyi(2009)]{rud09} Ruderman, M.~S., \& Erd{\'e}lyi, R.\ 2009, \ssr, 149, 199
\bibitem[Scherrer et al.(2012)]{sch12} Scherrer, P.~H., Schou, J., Bush, R.~I., et al.\ 2012, \solphys, 275, 207
\bibitem[Sharma(2017)]{sha17} Sharma, S.\ 2017, \araa, 55, 213
\bibitem[Shen et al.(2014)]{shen14} Shen, Y., Liu, Y.~D., Chen, P.~F., et al.\ 2014, \apj, 795, 130
\bibitem[Shen et al.(2017)]{shen17} Shen, Y., Liu, Y., Tian, Z., et al.\ 2017, \apj, 851, 101
\bibitem[Terradas et al.(2005)]{ter05} Terradas, J., Oliver, R., \& Ballester, J.~L.\ 2005, \aap, 441, 371
\bibitem[Terradas et al.(2007)]{ter07} Terradas, J., Andries, J., \& Goossens, M.\ 2007, \aap, 469, 1135
\bibitem[Tothova et al.(2011)]{toth11} Tothova, D., Innes, D.~E., \& Stenborg, G.\ 2011, \aap, 528, L12
\bibitem[Tripathi et al.(2009)]{tri09} Tripathi, D., Isobe, H., \& Jain, R.\ 2009, \ssr, 149, 283
\bibitem[Van Doorsselaere et al.(2008)]{van08} Van Doorsselaere, T., Nakariakov, V.~M., Young, P.~R., et al.\ 2008, \aap, 487, L17
\bibitem[Verwichte et al.(2004)]{ver04} Verwichte, E., Nakariakov, V.~M., Ofman, L., et al.\ 2004, \solphys, 223, 77
\bibitem[Vr{\v{s}}nak et al.(2007)]{vrs07} Vr{\v{s}}nak, B., Veronig, A.~M., Thalmann, J.~K., et al.\ 2007, \aap, 471, 295
\bibitem[Wang \& Solanki(2004)]{wang04} Wang, T.~J., \& Solanki, S.~K.\ 2004, \aap, 421, L33
\bibitem[White \& Verwichte(2012)]{wv12} White, R.~S., \& Verwichte, E.\ 2012, \aap, 537, A49
\bibitem[White et al.(2012)]{wht12} White, R.~S., Verwichte, E., \& Foullon, C.\ 2012, \aap, 545, A129
\bibitem[Yuan \& Van Doorsselaere(2016)]{yuan16} Yuan, D., \& Van Doorsselaere, T.\ 2016, \apjs, 223, 24
\bibitem[Zhang et al.(2014)]{zj14} Zhang, J., Li, T., \& Yang, S.\ 2014, \apjl, 782, L27
\bibitem[Zhang et al.(2012)]{zqm12} Zhang, Q.~M., Chen, P.~F., Xia, C., \& Keppens, R.\ 2012, \aap, 542, A52
\bibitem[Zhang et al.(2013)]{zqm13} Zhang, Q.~M., Chen, P.~F., Xia, C., et al.\ 2013, \aap, 554, A124
\bibitem[Zhang et al.(2016)]{zqm16} Zhang, Q.~M., Li, D., \& Ning, Z.~J.\ 2016, \apj, 832, 65
\bibitem[Zhang et al.(2017a)]{zqm17a} Zhang, Q.~M., Li, T., Zheng, R.~S., et al.\ 2017a, \apj, 842, 27
\bibitem[Zhang et al.(2017b)]{zqm17b} Zhang, Q.~M., Li, D., \& Ning, Z.~J.\ 2017b, \apj, 851, 47
\bibitem[Zhang \& Ji(2018)]{zqm18} Zhang, Q.~M., \& Ji, H.~S.\ 2018, \apj, 860, 113
\bibitem[Zhang et al.(2020a)]{zqm20a} Zhang, Q.~M., Guo, J.~H., Tam, K.~V., et al.\ 2020a, \aap, 635, A132
\bibitem[Zhang et al.(2020b)]{zqm20b} Zhang, Q.~M., Dai, J., Xu, Z., et al.\ 2020b, \aap, 638, A32
\bibitem[Zhou et al.(2018)]{zhou18} Zhou, Y.-H., Xia, C., Keppens, R., et al.\ 2018, \apj, 856, 179
\bibitem[Zimovets \& Nakariakov(2015)]{zim15} Zimovets, I.~V., \& Nakariakov, V.~M.\ 2015, \aap, 577, A4
\end{thebibliography}
\end{document}